\newif\ifnjp
\njpfalse

\ifnjp
\documentclass[12pt]{iopart} 
\expandafter\let\csname equation*\endcsname\relax
\expandafter\let\csname endequation*\endcsname\relax
\usepackage{amsmath}
\usepackage{iopams}
\else
\documentclass[10pt,pra,superscriptaddress,amsmath,amssymb,twocolumn]{revtex4-1}\usepackage{amssymb}  
\newcommand{\tr}{{\rm Tr}}  
\fi

\usepackage{amsthm}
\usepackage{graphicx}
\usepackage[all]{xy}
\usepackage[usenames,dvipsnames]{color}
\usepackage{tabularx}

\usepackage{amsmath,environ}  

\NewEnviron{prlequ}{%
\begin{equation}
  \BODY
\end{equation}
}

\renewcommand{\a}{f}
\renewcommand{\b}{g}

\newcommand{\mc}[1]{\mathcal #1}

\newcommand{\ave}[1]{\langle #1 \rangle}
\newcommand{\bra}[1]{\langle #1 |}
\newcommand{\ket}[1]{| #1 \rangle}
\newcommand{\braket}[2]{\langle #1 | #2 \rangle}

\newcommand{\proj}[1]{\ket{#1}\bra{#1}}
\newcommand{\one}{{\bf 1}}
\newcommand{\re}{{\rm Re}\,}

\theoremstyle{plain}
\newtheorem{proposition}{Proposition}

\theoremstyle{definition}

\begin{document}

\title{Inferring effective field observables from a discrete model}
\ifnjp
\author{C\'edric B\'eny$^{1,2}$}
\address{$^1$Institut f\"ur Theoretische Physik, Leibniz Universit\"at Hannover, Appelstra{\ss}e 2, 30167 Hannover, Germany} 
\address{$^2$Department of Applied Mathematics, Hanyang University (ERICA), 55 Hanyangdaehak-ro, Ansan, Gyeonggi-do, 426-791, Korea.}
\else
\author{C\'edric B\'eny}
\address{Institut f\"ur Theoretische Physik, Leibniz Universit\"at Hannover, Appelstra{\ss}e 2, 30167 Hannover, Germany} 
\address{Department of Applied Mathematics, Hanyang University (ERICA), 55 Hanyangdaehak-ro, Ansan, Gyeonggi-do, 426-791, Korea.}
\fi
\date{November 30, 2016} 

\begin{abstract}
A spin system on a lattice can usually be modeled at large scales by an effective quantum field theory. A key mathematical result relating the two descriptions is the quantum central limit theorem, which shows that certain spin observables satisfy an algebra of bosonic fields under certain conditions. Here, we show that these particular observables and conditions are the relevant ones for an observer with certain limited abilities to resolve spatial locations as well as spin values. This is shown by computing the asymptotic behaviour of a quantum Fisher information metric as function of the resolution parameters. The relevant observables characterise the state perturbations whose distinguishability does not decay too fast as a function of spatial or spin resolution.
\end{abstract}

\maketitle

Many interesting physical properties of solid materials can be modelled by spin systems, namely regular networks of finite-dimensional quantum systems which interact locally. Near a second order phase transition, the spins typically display collective behaviours which can be modelled by a quantum field theory (QFT). 
Given that the spin description underlies that in terms of fields, the field observables must have a precise representation as spin observables.

The quantum central limit theorem and its variations~\cite{cushen71,goderis89,ohya04} show that certain spin observables (the {\em fluctuation operators}) satisfy the same algebra as bosons in the thermodynamic limit (i.e., that of infinitely many spins). This statement holds weakly in terms of expectation values with respect to a product state. It can be extended so as to apply to a larger set of states, as well as to locally varying fluctuation operators~\cite{michoel98}. 

Here, we show that the form of these special observables can be derived from operational considerations, independent of the role they play in a central limit theorem. Moreover, our derivation provides a justification for the way the convergence is formulated. This provides a microscopic justification for the role that $n$-point functions play in quantum field theory, and establishes a systematic connection between the spin and the field description of a system.

Our derivation follows from answering the following question: which perturbations of a given state are most easily detectable provided certain limitations on experimental resolutions? 
We answer using the framework proposed in Refs.~\cite{beny13,beny14}. 

\section{Framework}

The approach relies on two inputs: a coarse-graining operation on states $\mc N$ (which we take to be a quantum channel, or completely positive trace-preserving map acting on density matrices~\cite{beny12}) depending on a family $r = (\sigma,y,\dots)$ of resolution parameters, and a distinguishability (Riemannian) metric on the manifold of density matrices (states). The metric is characterised by an inner product $\ave{\cdot,\cdot}_\rho$ on the tangent space at state $\rho$. The tangent space can be identified with the set of traceless self-adjoint operators as follows: if $f$ is a scalar function on states, then the operator $X$ is associated with the tangent vector $\tilde X$ satisfying $f(\rho + \epsilon X) = f(\rho) + \epsilon (\tilde Xf)(\rho) + \mc O(\epsilon^2)$. That is, $\tilde X$ is the derivative in the direction specified by $X$. 

The channel $\mc N$ transforms a tangent vector $X$ into $\mc N(X)$ (it is its own pushforward since it is linear.)
Hence it defines the coarse-grained metric $\ave{\mc N(\cdot),\mc N(\cdot)}_{\mc N(\rho)}$. We can interpret the coarse-grained distance 
\[
\|X- Y\|_{\mc N}^2 = \ave{\mc N(X-Y),\mc N(X-Y)}_{\mc N(\rho)}
\]
as a measure of distinguishability between $\rho + \epsilon X$ and $\rho + \epsilon Y$ to order $\epsilon^2$, for an observer with experimental resolutions specified by the family of parameters $r$ of $\mc N$. The definition of this manifold and information metrics can be done also for infinite-dimensional Hilbert spaces~\cite{kostecki16}.

We are interested in the amount by which a vector $X$ contracts under the coarse-graining, i.e., in the {\em contraction ratio} 
\begin{prlequ}
\tilde\eta(X) := {\|X\|_{\mc N}}/{\|X\|},
\end{prlequ}
where $\|X\| \equiv \|X\|_{\rm id}$.
Specifically, we want to characterise the asymptotic behaviour of $\tilde\eta(X)$ for large (coarse) resolutions parameters $r$. For instance, if $\tilde\eta(X)$ is zero, or decays exponentially with some components of $r$, then we can essentially ignore the tangent direction $X$ at $\rho$, as it is effectively unobservable. We want to characterise the real Hilbert space spanned by the remaining tangent vectors.

In order to classify subspaces by their contraction ratio, one could first maximize $\tilde \eta(X)$ over $X$ to find the least contracting (most relevant) vector $X_1$, then perform the maximization again in the complement of $X_1$ to find the next most relevant vector $X_2$, etc.
Mathematically, this is equivalent to solving the eigenvalue problem
\begin{equation} 
\label{evs}
\mc N_\rho^* \mc N (X_j) = \tilde \eta_j^2 X_j,
\end{equation}
where $\mc N_\rho^*$ is the adjoint of $\mc N$ with respect to the inner product $\ave{\cdot,\cdot}_\rho$. The eigenvectors $X_j$ are the principal directions of contraction, with respective contraction ratios $\tilde \eta_j$. 

Explicitely, $\mc N_\rho^*$ is defined by the relation
\begin{prlequ}
\ave{\mc N(X),Y}_{\mc N(\rho)} = \ave{X, \mc N^*_\rho(Y)}_\rho
\end{prlequ}
for all $X$ and $Y$.
For instance, for the $\chi^2$ metric $\ave{X,Y}_\rho = \tr(X \sqrt \rho \, Y \sqrt \rho)$, $\mc N^*_\rho$ is the channel introduced by Petz as the {\em transpose channel}~\cite{petz84}), which also plays a central role as approximate reversal of $\mc N$~\cite{barnum02,junge16}.

For a generic metric, which can be written as
\begin{prlequ}
\ave{X,Y}_\rho = \tr(X \Omega_{\rho}^{-1}(Y)),
\end{prlequ}
where $\Omega_\rho$ is a linear operator on density matrices, and $\Omega_\rho^{-1}$ its inverse which should be thought of as a representation of the metric as linear operator (kernel). The adjoint map is explicitely given by the composition
\begin{prlequ}
\mc N^*_\rho = \Omega_\rho \circ \mc N^\dagger \circ \Omega_{\mc N(\rho)}^{-1}.
\end{prlequ}
Here $\dagger$ denotes the adjoint with respect to the Hilbert-Schmidt inner product, i.e., $\tr(A \mc M(X)) = \tr(\mc M^\dagger(A) X)$ for all operators $A$, $X$. This is the Heisenberg-picture representation of the channel $\mc N$. Notice that time flows ``backward'' in that picture, as $(\mc M \mc N)^\dagger = \mc N^\dagger \mc M^\dagger$.

Here, for simplicity, we focus exclusively on the Bures metric given by
\begin{prlequ}
\Omega_\rho(A) = \frac 1 2 \left({ \rho A + A \rho }\right),
\end{prlequ}
which is the smallest of the contractive metrics (when normalized to match the Fisher metric on diagonal density matrices)~\cite{petz96,petz96b}. This metric is well defined on the submanifold of pure states.

We do not need an explicit expression for $\Omega_\rho^{-1}$. Instead, we consider tangent vectors of the form $X = \Omega_\rho(A)$ and $Y = \Omega_\rho(B)$, so that $\ave{X,Y}_\rho = \tr(A \,\Omega_\rho(B)) = \re \tr(\rho A B)$. For convenience, we write
\begin{equation}
\eta(A) := \tilde \eta(X).
\end{equation}
It is useful to think of $A$ as representing the cotangent vector image of the tangent vector $X$ by contraction with the metric. 
The traceless condition on $X$ becomes $\tr(\Omega_\rho(A)) = \tr(\rho A) = 0$. 
Hence we represent cotangent vectors at $\rho$ by self-adjoint operators of zero expectation value with respect to $\rho$. 
We observe also that, as $X$ is mapped to $\mc N(X)$, its cotangent representation $A$ is mapped to $(\mc N^*_\rho)^\dagger(A)$.
Indeed, one can directly check that
\begin{prlequ}
\mc N(X) = \Omega_{\mc N(\rho)}\bigl((\mc N^*_\rho)^\dagger(A)\bigr).
\end{prlequ}
In fact the role of $\mc N$ and $\mc N^*_\rho$ is reversed as, assuming $Y = \Omega_{\mc N(\rho)}(B)$, then
\begin{prlequ}
\mc N_\rho^*(Y) = \Omega_\rho( \mc N^\dagger(B)).
\end{prlequ}

On the boundary of the manifold of states, corresponding to those states $\rho$ which are not invertible, there are directions $X$ which cannot be written as $\Omega_\rho(A)$. However, assuming $\mc N(\rho)$ is inside the bulk, hence invertible, those $X$ have $\tilde \eta(X) = 0$ and can therefore be neglected. Indeed, for such $X$, as $\rho$ tends to the boundary $\|X\| \rightarrow \infty$, but since $\|X\|_{\mc N}$ converges, $\tilde \eta(X)$ tends to zero. 
\begin{proposition}
\label{prop1}
If $\mc N(\rho)$ is invertible, then 
a tangent vector $X$ at $\rho$ is {\em irrelevant}, i.e., $\tilde \eta(X) = 0$, whenever it is not of the form $X = \Omega_\rho(A)$, where $A = A^\dagger$ is such that $\tr(\rho A^2) \neq 0$.
\end{proposition}

Said differently, the relevant tangent vectors all live in the {\em Gelfand-Naimark-Segal (GNS) representation} of the algebra of operators with respect to $\rho$. Invoking the GNS construction here may be somewhat extravagant as we are only considering finite-dimensional Hilbert spaces. However, this provides a compact way of thinking about the above proposition, and it suggests a natural C$^*$-algebraic generalisation of our approach.

The GNS construction works as follows.
Let us consider the complex Hilbert space $\mc H$ (resp.~$\mc H'$) whose vectors are labelled by operators $A$, equipped with the inner product $\braket A B := \tr(\rho A^\dagger B)$ (resp.~$\braket A B' := \tr(\mc N(\rho) A^\dagger B)$). Then, provided $\mc N(\rho)$ is invertible, we can define the linear operator $N: \mc H' \rightarrow \mc H$ which maps $\ket A$ to $\ket{\mc N^\dagger(A)}$ for any operator $A$. As noted above, this is in fact the cotangent representation of the action of $\mc N^*_\rho$.
Operators also naturally act on $\mc H$ as $A\ket B := \ket{AB}$. 
  
The linear map $N^*: \mc H \rightarrow \mc H'$ representing the cotangent action of the original map $\mc N$ is {\em not} given by the complex adjoint of $N$. It is instead defined via the real inner product through $\re \bra B N^* \ket A' = \re \bra A N \ket B$. 
In this formalism, Equ.~\eqref{evs} becomes
\begin{equation}
\label{evh}
N N^* \ket A = \eta^2 \ket A,
\end{equation} 
and the contraction ratio of an arbitrary operator $A$ is given by 
\begin{equation}
\eta(A)^2 = \frac{\bra A N N^*\ket A }{ \braket A A }.
\end{equation} 

This could be generalised in principle to a setting where the state $\rho$ is a positive linear functional on a C$^*$-algebra $\mc A$, and the channels $\mc N^\dagger$ are completely positive unital maps from some algebra $\mc A'$ to $\mc A$, provided that the states $\rho \circ \mc N^\dagger$ are faithful.
In this representation, the relevant part of the tangent space at $\rho$ is then the real subspace of $\mc H$ spanned by $\ket A$ where $A^\dagger = A$ and $\rho(A) = 0$, with metric given by $\re \braket A B$.

\section{Spatially homogeneous case}
In what follows, we consider $n$ quantum systems, which we refer to as {\em spins} or {\em sites}, each of dimension $d$.
For the first example, we assume that the observer cannot choose which spin they address, and that, when measuring a spin, they do so with resolution $y$. This can be formalised by assuming that they only have access to the coarse-grained states $\mc N(\rho + \epsilon X)$ where $\mc N = \mc P \circ \mc D^{\otimes n}$. The projective map $\mc P(\rho) = \frac 1 {n!} \sum_\pi U_\pi \rho U_\pi^\dagger$ is the average over all permutations of the $n$ spins, where $U_\pi$ is the unitary operator implementing the permutation $\pi$. The channel $\mc D^{\otimes n}$ is the parallel application of the depolarization map 
\begin{equation}
\label{depol}
\mc D(\rho) := \tfrac 1 y \rho + \bigl( 1 - \tfrac 1 y \bigr) \tr(\rho) \tfrac \one d,
\end{equation}
to each spin, where $y \in [1,\infty)$ represents the resolution, or {\em imprecision}, at which spin measurements are resolved. Observe that $\mc D^{\otimes n}$ and $\mc P$ commute, and that $\mc D^\dagger = \mc D$ and $\mc P^\dagger = \mc P$.
In this example, the channel $\mc N$ possesses a single resolution parameter $y$ (hence $r = y$).

We consider the case of a product state $\rho^{\otimes n} = \rho \otimes \dots \otimes \rho$. 
As explained above (Proposition~1), the relevant part of the tangent space at $\rho^{\otimes n}$ can be represented by vectors $\ket A$ in the GNS representation with respect to $\rho^{\otimes n}$. 

Let us now show that, due to the depolarisation maps, we only need to consider vectors $\ket A$ where $A$ is $k$-local for any finite $k$ independent of $n$, because the contraction ratio $\eta(A)$ for any non-$k$-local operator $A$ (to be defined below) is bounded by a function of $k$ which tends to zero as $k$ goes to infinity.

To formulate this more precisely,
consider the space $V$ (resp.~$V'$) of single-site operators $\a$ such that $\tr(\rho \a) = 0$ (resp.~$\tr(\mc D(\rho) \a) = 0$), and let $\mc T_k$ (resp.~$\mc T_k'$) denotes the space spanned by the operators of the form
 $\a_1^{(i_1)} \cdots \a_j^{(i_j)}$, $0 \le j \le k$, where $\a_i \in V$ (resp.~$\a_i \in V'$), and $\a^{(i)}$ denotes the operator $\a$ acting on site $i$. We also include $\one$ in $\mc T_k$. 

Since the information metric contracts under the action of any channel, we have $\|X\|_{\mc N} \le \|X\|_{\mc D^{\otimes n}}$. Hence, considering only the depolarisation map, one can show (see the Appendix) that, in this setting,
\begin{proposition}
\label{prop2}
$\eta(A) \le \mc O(y^{-(k+1)})$ for all $A$ such that $\re \braket A B = 0$ for all $B \in \mc T_k$. That is, for $A$ orthogonal to all $k$-local observables. 
\end{proposition}

Now let us consider the effect of the channel $\mc P$. This is simpler because $\mc P$ is projective, which implies that all tangent vectors represented by operators in its kernel are irrelevant. Accordingly, we can directly eliminate such vectors $\ket A$ where $A$ is not fully symmetric under permutation of the spins. Indeed, let $P$ such that $P \ket A = \ket{\mc P(A)}$. We have 
\begin{equation}
\begin{split}
\re \bra B P^* \ket A &= \re \tr(A \mc  P(B) \rho^{\otimes n}) = \re \tr(A \mc P(B \rho^{\otimes n})) \\
&= \re \tr(\mc P(A) B \rho^{\otimes n}) = \re \bra B P \ket A.
\end{split}
\end{equation}
Hence $P^* = P$. Since also $P^2 = P$, it is an orthogonal projector. It follows that all vectors $\ket A$ orthogonal to symmetric ones are such that $P \ket A = 0$, hence $N N^* \ket A = P D D^* P \ket A = 0$, where we wrote $D$ for the representation of $\mc D^{\otimes n}$. 

Combining this observation with Proposition~\ref{prop2}, we conclude that, for the state $\rho$ and channel $\mc N$ introduced above,
\begin{proposition}
\label{prop3}
The eigenspaces of $NN^*$ for eigenvalues up to order $y^{-k}$ are contained in the space spanned by $\ket{A}$ for any fully symmetric $A \in \mc T_k$, namely
\begin{equation}
\mc H_k^S := \{ \ket{\mc P(A)} : A \in \mc T_k \}.
\end{equation}
\end{proposition}
Hence, if we want to characterise the tangent directions with contraction ratio only down to order $y^{-k}$, we can restrict the analysis to the subspace generated by the $k$-local symmetric observables. Since this statement is independent of $n$, we can take the thermodynamic limit $n \rightarrow \infty$ with fixed $k$: this is the setting of the central limit theorem. 

We observe that the dimension of $\mc H_k^S$ does not depend on the number of sites $n$. Only the scalar product $\braket A B$ does. Accordingly, we formulate the central limit as a limit of a sequence of scalar product on a fixed vector space, which we take to be the complex vector space
\begin{equation}
\mc F_k := \bigoplus_{j=0}^k V^{\odot j},
\end{equation}
where $\odot$ denotes the symmetrised tensor product and $V^{\odot 0} = \mathbb C$ is the ``vacuum'' sector. Using the operators
\begin{prlequ}
G_\a = \bigl( \one + \tfrac i {\sqrt n}\a \bigr)^{\otimes n},
\end{prlequ}
we define the surjective linear map $\alpha: \mc F_k \rightarrow \mc H_k^S$ by
\begin{prlequ}
\alpha(\a_1\odot \dots\odot \a_j) = \partial_{t_1}\cdots \partial_{t_j} \ket{G_{-i\Sigma_i t_i \a_i}}|_{t_1 = \dots = t_j = 0}.
\end{prlequ}
For any $u \in \mc F_k$, we abbreviate the corresponding linear combination of differentiations as $\alpha(u) = \Delta_{\a}^u \ket{G_{-i\a}}$.
For instance, we get $\alpha(\a) = \ket{ F_\a}$, where $F_\a := \tfrac 1 {\sqrt n} \sum_i \a^{(i)}$ is a {\em fluctuation operator}. 
(Recall that $f^{(i)}$ denotes the operator $f$ acting on site $i$).
Also,
\begin{prlequ}
\alpha(\a \odot \b) = \ket{F_f F_g} - \tfrac 1 {\sqrt n} \ket{F_{\a \b }}.
\end{prlequ}
The vacuum is mapped to $\ket{\one}$.

The GNS inner product is represented on $\mc F_k$ as
\begin{prlequ}
{\braket u v}_n := \braket{\alpha(u)}{\alpha(v)} = \Delta_\a^u \Delta_\b^v \braket{G_{-i\a}}{G_{-i\b}},
\end{prlequ}
where we have explicitely 
\[
\braket{G_{-i\a}}{G_{-i\b}} = \braket{G_{\a}}{G_{\b}} = (1 + \tfrac 1 n \tr(\rho \a^\dagger \b))^n.
\]
 
The eigenvalue problem for $NN^*$ can be simplified greatly by working with the limiting metric 
\begin{prlequ}
\label{innerprod}
\braket{u}{v} = \lim_{n\rightarrow \infty} \braket{u}{v}_n = \Delta_\a^u \Delta_\b^v \;e^{\tr(\rho \a^\dagger \b)}
\end{prlequ}
on $\mc F_k$: this leads to a form of the quantum central limit theorem.
However, we need to make sure that this does not amount to cheating: i.e., that no vector that is relevant in terms of $\braket{\cdot}{\cdot}_n$ become artificially irrelevant in terms of $\braket{\cdot}{\cdot}$. In other word, that $\braket{u}{u} = 0$ implies $\braket{u}{u}_n = 0$ for all $u \in \mc F_k$.
This is indeed the case, which can be seen from the fact that the subspaces $V^{\odot i}$ and $V^{\odot j}$ for $i \neq j$ are orthogonal in both metrics, and within $V^{\odot j} \subseteq \mc F_k$ both metrics are proportional to each other. Indeed, both generators $\braket{G_{\a}}{G_{\b}}$ and $\braket{G_{\a}}{G_{\b}}_n$ are power series in $\tr(\rho\a^\dagger \b)$ with no zero coefficient.

From the explicit form of the limiting inner product (Equ.~\ref{innerprod}), one recognises that the completion of $\mc F_k$ with respect to it (that is, once zero norm vectors have been modded out) is the $0$-to-$k$-particle subspace of the symmetric Fock space $\hat{\mc F}$ built from the single particle space Hilbert space $V$ with the inner product $\tr(\rho \a^\dagger \b)$.  Indeed, let $a_\a$ denotes the annihilation operators on $\hat{\mc F}$ satisfying $[a_\a,a_\b^\dagger] = \tr(\rho \a^\dagger \b) \one$, and let $\phi(\a) = a_\a + a^\dagger_\a$. Consider the normally ordered displacement (Weyl) operators, also called {\em vertex operators},
\begin{equation}
\hat G_\a = e^{i a^\dagger_\a} e^{i a_\a} = e^{i \phi(\a)} e^{\frac 1{2} \tr(\rho \a^\dagger \a)} 
\end{equation}
on $\hat{\mc F}$.
Then one can check that, if we write $\Omega$ for the vacuum in $\hat{\mc F}$, then
\begin{equation}  
\bra{\Omega}\hat G_\a^\dagger \hat G_\b \ket{\Omega} = e^{\tr(\rho \a^\dagger \b)} = \lim_{n\rightarrow \infty} \braket{G_\a}{G_\b}.
\end{equation}
Hence the vacuum plays the role of our reference state $\rho^{\otimes n}$, and $\alpha(u) \in \mc H_k^S$ is represented by $\Delta_f^u \hat G_f \in \hat{\mc F}$.

The same construction can be done with respect to the state $\mc D(\rho)$ instead of $\rho$, yielding the spaces $\mc F_k'$, limiting Fock space $\hat{\mc F}'$ and vertex operators $\hat G_f'$ for $f \in V'$.

Noting that $N \ket{G_\a} = \ket{G_{\mc D(\a)}}$ for $\a \in V'$, we see that the map $N$ is represented on $\hat{\mc F}'$ by $\hat N\ket{\hat G_\a'} = \ket{\hat G_{\mc D(\a)}}$, which is the tangent action of the gaussian channel $\hat{\mc N}$ defined by $\hat{\mc N}^\dagger(\hat G_{\a}) = \hat G_{\mc D(\a)}$. This a form of central limit for channels, as done in Ref.~\cite{quaegebeur84}.

These results allow us to fully solve the eigenvalue problem corresponding to Equ.~\eqref{evs} within the Gaussian formalism using the method introduced in Ref.~\cite{beny15}, for the channel $\hat{\mc N}$ and at the tangent space to the vaccuum state.
We find that $\hat N \hat N^*$ is block-diagonal, where each block corresponds to a given order of polynomial generated by $\hat G_\a$. For order $k$ polynomials, this is an eigenvalue problem in a vector space of dimension $(d^2-1)^k$. Also we know that the polynomials of order $k$ or larger have contraction ratio of order $\mc O(y^{-k})$.

\subsection{Example}

For instance, consider the case where the dimensionality of each site is $d=2$, each in a pure state $\rho = \proj 0$. Let $\tau_i$, $i \in \{1,2,3\}$ denote the Pauli matrices in the basis $\ket 0$, $\ket 1$. A convenient basis of the cotangent space $V$ at one site (zero expectation value self-adjoint operators) consists of the matrices $f_1 = \tau_1$, $f_2 = \tau_2$ and $f_3 = \tau_3 - \one$. 
Similarly, a basis of the cotangent space $V'$ at $\mc N(\rho)$ is given by $f_1' = \sqrt {y} \,\tau_1, f_2' = \sqrt {y} \,\tau_2$ and $f_3' = \tau_3 - \one/y$, where recall that $y$ is the depolarization parameter.

As shown in the general case, in the limit $n \rightarrow \infty$, we can study the effect of the channel on the tangent space at $\rho^{\otimes n}$ by replacing our system with a family of Fock spaces $\hat{\mc F}'$ parameterized by $y$, where the vacuum plays the role of the state $\mc D(\rho^{\otimes n})$ and the channel $\mc N$ corresponds to the Gaussian channel defined by $\hat{\mc N}^\dagger(\hat G_{\a}') = \hat G_{\mc D(\a)}$ where $G_f'$ are normally ordered Weyl operators and $G_f$ is $G_f'$ for $y=1$.

For any $y$, $\hat{\mc F}'$ is built by second quantization of the Hilbert space given by $V'$ equipped with the form
\begin{equation}
\label{exprod}
\braket{f_i'}{f_j'} = \tr(\mc D(\proj 0) f_i' f_j') = y\delta_{ij} + i \Delta_{ij}
\end{equation}
for $i,j=1,2$ and zero for all other components. The matrix $\Delta_{ij}$ is antisymmetric with $\Delta_{12} = 1$.
Since the norm of $f_3'$ is zero, we must eliminate this vector, so that we are left with a two-dimensional Hilbert space spanned by $f_1'$ and $f_2'$.

The imaginary part of this expression gives us the commutator for the quantized version of the elements of $V'$.
Let us write $\hat x := \phi'(f_1')/\sqrt 2$ and $\hat p := \phi'(f_2')/\sqrt 2$. We see that these operators satisfy the canonical commutation relations $[\hat x,\hat p] = i \one$.
Since these commutation relations are independent of the spin precision $y$, we can use the same CCR algebra for all $y$, including $y=1$. 

In the basis $f_1', f_2'$, the covariance form for the vacuum state is represented by the 2-by-2 matrix $y \one$ (real component of Equ.~\eqref{exprod}), and the gaussian channel $\hat{\mc N}^\dagger$ maps a gaussian state with covarience matrix $M$ to one with covarience matrix $\tfrac 1 {y} M  + (y - \tfrac 1 {y}) \one$. 

We these results we can now proceed as in Ref.~\cite{beny15} and solve the eigenvalue problem of Eq.~\eqref{evs} in details. 
The space of polynomials of degree $k$ generated by $\hat G_f'$ can be parameterized using the $k$-fold tensor products of the basis $\{f_1',f_2'\}$ on the space $(V')^{\otimes k}$. The coarse-grained metric has components compiled in the matrix $K_y^k = \re \bigl( (K_y)^{\otimes k} \bigr)$ where $(K_y)_{ij} = y\delta_{ij} + i \Delta_{ij}$. Hence, the pure metric has components given by $K_1^k$.
Moreover, the components of $\hat N$ are given by the matrix $y^{-k/2} \one$, which can be seen from the fact that $\mc D^\dagger(f_{1,2}') = y^{-\frac 1 2} f_{1,2}$. The components of the linear map $\hat N \hat N^*$ to be diagonalised are given by $y^{-k} (K_y^k)^{-1} K_1^k$. For instance, for $k=1$, this yields the eigenvectors $\hat x$ and $\hat p$ both with eigenvalue $y^{-2}$, then for $k=2$, $\hat x^2-\hat p^2$ and $\hat x \hat p + \hat p \hat x$ have eigenvalue $\tfrac{2 y^{-2}}{1+y^2}$ and $\hat x^2 + \hat p^2$ and $i(\hat x \hat p - \hat p \hat x)$ have eigenvalue zero, etc. In turn, one can find the direct spin representation of the corresponding tangent vectors. For instance, $\hat x \hat p + \hat p \hat x$ corresponds to $\tfrac 1 2 \alpha(f_1 \odot f_2) = \tfrac 1 {2n} \sum_{i\neq j} \tau_1^{(i)} \tau_2^{(j)} \ket I$.

\section{Quantum field theory}
The previous result is the spatially homogeneous, or ``0-mode'' version of a more general situation where a true quantum field theory with local degrees of freedom emerges.  

We consider again $n$ independent $d$-dimensional quantum systems, but this time assume that they are arranged on a regular spatial lattice in $D$ dimensions of space with lattice spacing $\epsilon$.

We use the coarse-graining introduced in Ref.~\cite{beny12}, namely
\(
\mc N = \mc P \circ \mc D^{\otimes n},
\)
where the single-site channel $\mc D$ is defined in Equ.~\eqref{depol} and
\(
\mc P =  e^{\frac 1 2 (\sigma/\epsilon)^2 \mc L}
\)
with the generator
\begin{equation}
\mc L(A) = \sum_{\ave{ij}} (U_{ij} A U_{ij}^\dagger - A)
\end{equation}
where the sum runs over neighbouring sites $i$ and $j$ and $U_{ij} = U_{ij}^\dagger$ unitarily swaps them.
It is easy to see that $\mc D^{\otimes n}$ commutes with $\mc P$ and both are self-adjoint with respect to the Hilbert-Schmidt inner product. 
Here the resolution parameters of $\mc N$ are $r = (\sigma,y)$. As before, $y$ represents a local spin resolution, whereas $\sigma$ is now a spatial resolution, with unit of distance. 

The map $\mc P$ is well defined on an infinite lattice, as a map on the {\em quasilocal algebra} $\mc A$, i.e., the C$^*$-algebra generated by local operators. This allows us to work directly in the thermodynamic limit $n \rightarrow \infty$. Instead, this time we are concerned with the {\em continuum limit} $\epsilon \rightarrow 0$. 

The intuition is the following: if we focus on a region of size $L \ll \sigma$, then $\mc P$ fully symmetrises the lattice in that region, hence acting as the channel that we used in the previous example. As $\epsilon \rightarrow 0$ the number of sites in that region is of order $(L/\epsilon)^D$. Hence, locally, the continuum limit $\epsilon \rightarrow 0$ looks just like the limit $n \rightarrow \infty$ in the previous calculation. 

We work with the product state $\omega \equiv \rho^{\otimes \infty}$, which is a well defined state on $\mc A$.
As in the previous example, the action of the local depolarization channels implies that $\eta(A) \le \mc O(y^{-(k+1)})$ whenever $A$ is orthogonal to all $k$-local operators. This allows us to also work within the part of the tangent space corresponding to $k$-local operators $\mc T_k$ and $\mc T_k'$.

Since $\mc P$ implements a convex combination of permutations of lattice sites, $\omega$ is a fixed point of $\mc P$. If $P$ denote the GNS representation of $\mc P$ with respect to $\omega$, then it follows that $P^* = P$, and hence
\begin{equation}
\eta^2(A) \le \frac{\bra{A}PP^*\ket{A}}{\braket{A}{A}} = \frac{\bra{A}P^2\ket{A}}{\braket{A}{A}}.
\end{equation} 
On single site operators in $\mc T_1$, $\mc L$ generates a diffusion on the lattice~\cite{beny12}. It follows that, if we denote by $\mc T_1^\Lambda$ the set of operators of the form $A = \sum_{j \in \mathbb Z^D} f(j \epsilon )^{(j)}$ where $f: \mathbb R^D \rightarrow \mc V$ is bandlimited, i.e., its Fourier transform is supported in the ball $|p| < \Lambda$, then, for all $A \in (\mc T_1^{\Lambda})^\perp$, 
\begin{equation}
\eta(A) \le \mc O( e^{-\frac 1 2 \sigma^2 \Lambda^2 } ).
\end{equation}

On products of $k$ single site operators (which span $\mc T_k$), $\mc L$ generates a permutation of the $k$ sites. 
Leaving a proof for future work, let us here argue that this permutation corresponds to a sufficiently good approximation to $k$ independent random walks on the lattice, independently of the dimension $D$ of space. Indeed, deviations from this approximation happen when two walkers would find themselves on the exact same site according to independent random walks. Here, instead, the walkers swap positions and hence just pass each other, causing one of the walker's position to be shifted by one site relative to independent walkers. 
In one dimension, for instance, this implies that the difference is always at most a shift by $k$ sites for each walker, since two crossing by the same two walkers can only undo the shift caused by the first crossing~\footnote{This argument was suggested to the author by David Gross.}. This is true also in higher dimension along each components.

Therefore, we expect that for large $\sigma$ the effect of the map $\mc P$ on $\mc T_k$ does not differ significantly from that of $k$ independent diffusions on the lattice.
Hence, only momentum components $p < 1/\sigma$ should be relevant also for operators in $\mc T_k$. 

In order to formalize this statement, we need to introduce a few tools.
Given a function $f$ assigning a single site operator $f(x) \in V$ to each $x \in \mathbb R^D$, we introduce the operators
\begin{prlequ}
\label{qftgen}
G_f := \prod_{j \in \mathbb Z^D}\bigl( \one + i \epsilon^{D/2} f(j \epsilon)^{(j)} \bigr).
\end{prlequ}
The convergence of this product inside $\mathcal A$ may require $f$ to decrease fast enough spatially. However, derivatives of $G_f$ of finite order at $f=0$ such as $\Delta^u_f G_{-if}$ are well-defined for any variations $u$.

Let us denote by $\mc V_\Lambda$ the space of band-limited functions from $\mathbb R^D$ to $V$ with cutoff $\Lambda$, i.e., functions which are Fourier transforms of functions supported on the ball of radius $\Lambda$.
As in the homogeneous case, we consider the vector space
\begin{equation}
\mc F_k^\Lambda := \bigoplus_{j=0}^k \mc V_\Lambda^{\odot j}
\end{equation}
and define the surjective linear map
\(
\alpha: \mc F_k^\Lambda  \rightarrow \mc T_k,
\) 
via \(
\alpha(u) = \Delta_f^u G_{-if}.
\)

Let us write
\begin{equation}
\mc T_k^\Lambda := \alpha(\mc F_k^\Lambda)
\end{equation}
for the image of $\alpha$.
Our statement then, is that for all $A \in (\mc T_k^\Lambda)^\perp$, 
we have
\begin{equation}
\label{unproven1}
\eta(A) \le \mc O(e^{-\tfrac 1 2 k \,\sigma^2 \Lambda^2}).
\end{equation}
That is, components of rapid spatial variations are essentially irrelevant, as long as $\Lambda > 1/\sigma$. Importantly, this expression does not depend on the lattice spacing $\epsilon$.

We now use this fact to obtain a continuum limit of the tangent space without neglecting any potentially relevant vectors. We do so by noting that the subspace $\mc T_k^\Lambda$ (which contains all potentially relevant vectors) can be represented in $\mc F_k^\Lambda$ so that the dependance on $\epsilon$ is purely contained in the definition of the scalar product, namely

\begin{prlequ}
\braket{u}{v}_\epsilon = \braket{\Delta_f^u G_f}{\Delta_g^v G_g}.
\end{prlequ}
This can be computed using the fact that 
\begin{prlequ}
\braket{G_f}{G_g} = \prod_{j \in \mathbb Z^D}\bigl( 1 + \epsilon^D \tr(\rho f(j\epsilon)^\dagger g(j \epsilon)) \bigr).
\end{prlequ}
We then obtain the continuum limit by completing $\mc T_k^\Lambda$ with respect to
\begin{prlequ}
\braket{u}{v} := \lim_{\epsilon\rightarrow 0} \braket{u}{v}_\epsilon.
\end{prlequ}
Observing that
\begin{equation}
\lim_{\epsilon\rightarrow 0}\braket{G_f}{G_g} =  e^{\braket{f}{g}},
\end{equation}
where
\begin{equation}
\braket{f}{g} = \int_{\mathbb R^D} \tr(\rho f(x)^\dagger g(x)) \,dx
\end{equation}
is the value of the limiting inner product on the one particle sector, then
\begin{prlequ}
\braket{u}{v} = \Delta_f^u\Delta_g^v \, e^{\braket{f}{g}}
\end{prlequ}
which we recognize as the scalar product of the symmetric Fock space $\hat{\mc F^\Lambda}$ built from the single particle Hilbert space defined by $\mc V_\Lambda$ equipped with $\braket{f}{g}$. This is our continuum limit. 

We have to check again that we are not cheating. Namely that completing $\mc F_k^\Lambda$ with respect to $\braket{\cdot}{\cdot}$ does not eliminate any relevant vector, i.e., that $\lim_{n \rightarrow \infty} \braket{u_n}{u_n} = 0$ implies $\lim_{n \rightarrow \infty} \braket{u_n}{u_n}_\epsilon = 0$ for all $\epsilon > 0$. In the one particle sector, the discrete scalar product is just a discretization of the integral. The vanishing of the integral implies that of the sum because our functions are bandlimited: they cannot become arbitrarily peaked so as to converge to a function that is nonzero only at certain points. This argument can be extended to the multiparticle sectors.

In the limit, the operators $G_f$ play the same role as the normal ordered displacements (Weyl) operators
\begin{prlequ}
\hat G_f = e^{i \phi(f) + \frac 1 2 \braket{f}{f}}
\end{prlequ}
defined on the Fock space $\hat{\mc F^\Lambda}$,
where $\phi(\a) = a_f + a^\dagger_f$ and $a_f$ denotes annihilation operators, in the sense that, denoting by $\Omega$ the vacuum in $\hat{\mc F^\Lambda}$,
\begin{prlequ}
\lim_{\epsilon \rightarrow \infty} \braket{G_f}{G_g} = \bra{\Omega}\hat G_f^\dagger \hat G_g \ket{\Omega}.
\end{prlequ}

If, as argued above, the channel $\mc P$ factors for large $\sigma/\epsilon$ as independent diffusions on the lattice for each term of a product of single-site operators, then it acts in the limit as follows: 
\begin{prlequ}
\label{unproven2}
\lim_{\epsilon \rightarrow 0} \Delta_f^{u} \Delta_g^{v} 
\bra{G_f} P \ket{G_g} = \Delta_f^{u} \Delta_g^{v}\bra{\Omega}\hat G_f^\dagger \hat G_{X g} \ket{\Omega}, 
\end{prlequ}
for all $u,v \in \mc F_k$,
where the linear map $X : \mc V' \rightarrow \mc V$ is given by
\begin{prlequ}
(X f)(x) = \int_{\mathbb R^D} g_\sigma(x-x') f(x')\,dx',
\end{prlequ}
and $g_\sigma$ is a normalised gaussian of variance $\sigma$.

This implies that the map $N$ is represented in the continuum limit by the gaussian channel $\hat {\mc N}$ satisfying
\begin{prlequ}
\hat{\mc N}^\dagger(\hat G_f') = \hat G_{\mc D(X f)},
\end{prlequ}
where $\hat G_f'$ is built as $\hat G_f$ but using test functions taking value on $V'$ rather than $V$.

One could now finish solving the eigenvalue problem (Equ.~\eqref{evh}) in terms of this gaussian representation of the relevant tangent space and channel. We refer to Refs.~\cite{beny14} for a general analysis, and to Ref.~\cite{beny15} for a general solution method. The key is to recognise that the map $N N^*$ is block diagonal with respects to families of $k$ modes $p_1,\dots,p_k$. Each of these block is finite dimensional, allowing for a per order solution. Moreover, for any $A$ in such a block, $\eta(A) \le \mc O(y^{-k} e^{-\frac 1 2 \sum_i p_i^2 \sigma^2})$ in terms of the resolutions $y$ and $\sigma$.

\section{Discussion}
We have argued that, among all infinitesimal perturbations of the product state of a spin system, which includes arbitrarily correlated ones, only product of slowly varying fluctuations operators are distinguishable given certain reasonable experimental limitations. This justifies the application of a continuum limit taking the form of a central limit by which the relevant perturbations are identified with a subset of the perturbations of a quasi-free bosonic field state. 

In the inhomogeneous case, what is missing to obtain a fully rigorous argument are proofs for Eq.~\eqref{unproven1} and Eq.~\eqref{unproven2}, which depend on a bound characterizing the difference between the effect of the channel $\mc P$ on products of local operators and the product of the image of these local operators under $\mc P$.

This approach possesses some remarkable traits which require further analysis:
\begin{enumerate}
\item
The Hilbert space of the emergent quantum field theory corresponds formally to a tangent space of an underlying microscopic state. This formalises the intuition that the effective quantum field theory describes quasiparticles which are linear perturbation of an equilibrium state.
\item
The information metric and coarse-graining quantum channel play an explicit role in  identifying the relevant vectors. The algebra of the emergent quantum field theory is hence determined by a subtle interplay between the intrinsic correlation properties of the base state, and these {\em extrinsic} aspects of the observational setting. 
\item
Proposition~\eqref{prop1} provides an interesting physical interpretation for the GNS construction, provided it is relevant in infinite-dimensional settings.
\item
This approach provides a direct connection between a real-space renormalisation scheme on a lattice, represented by the family of quantum channels, to the standard momentum space renormalisation group of effective quantum field theories (see Ref.~\cite{beny14} for more details). 
\end{enumerate}

Altogether, these results validate an approach which allows in principle for the systematic derivation of relevant fluctuation observables around a given state, and for a given experimental situation. It will be most interesting to apply it to critical systems, or highly correlated spin states such as the toric code~\cite{kitaev03}, where the tangent vectors are labelled by string-like operators and the continuum limit is expected to be a topological quantum field theory~\cite{balsam12}.

\section{Acknowledgements}
The author is grateful to Tobias Osborne for discussions leading to this work.
This work was supported by the ERC grants QFTCMPS and SIQS, by the cluster of excellence EXC 201 Quantum Engineering and Space-Time Research, and by the research fund of Hanyang University (HY-2016-2237).

\ifnjp
\section*{References}
\bibliographystyle{iopart-num}
\fi

\bibliography{product_njp}

\newpage

\section*{Appendix: Irrelevance of non-local operators}

Let $\braket A B := \tr({\rho^{\otimes n}} A^\dagger B)$.
Given a finite set of sites $\Sigma$, let $\mc T_{\Sigma}$ denote the linear space spanned by the operators of the form $\prod_{i \in \Sigma} A_i^{(i)}$ such that $\tr(\rho A_i) = 0$ for all $i$, where $A^{(i)}$ denotes the single site operator $A$ acting on site $i$.
The space of $k$-local operators is 
\begin{equation}
\mc T_k := \bigoplus_{|\Sigma| \le  k} \mc T_\Sigma,
\end{equation}
which includes the identity operator.
We denote by $\mc T_k^\perp$ the space of operators orthogonal to $\mc T_k$ in terms of $\braket{\cdot}{\cdot}$. 
It is easy to see that $\mc T_\Sigma$ and $\mc T_{\Sigma'}$ are orthogonal whenever $\Sigma \neq \Sigma'$. Since the spaces $\mc T_\Sigma$ also span all operators, we have
\begin{equation}
\mc T_k^\perp = \bigoplus_{|\Sigma| > k} \mc T_\Sigma.
\end{equation}
The self-adjoint parts of the spaces $\mc T_\Sigma$ are also orthogonal for different $\Sigma$'s in terms of the metric, which is given by $\re\braket \cdot \cdot$.

\begin{proposition}
\label{theprop}
Let $A \in \mc T_{k-1}^\perp$, $A = A^\dagger$ and $X = \Omega_{\rho^{\otimes n}}(A)$, then
\begin{equation}
\label{yscaling} 
\|X\|_{\mc D^{\otimes n}}^2 \le \frac{d^k \,y^{-2k} }{(1-y^{-1})^k} \|X\|^2
\end{equation}
provided that $y(y-1) > d$.
\end{proposition}

\proof
Below, we write $\mc N := \mc D^{\otimes n}$.
The coarse-grained norm of a tangent vector represented by $X = X^\dagger$ is
\[
\begin{split}
\|X\|_{\mc N}^2 = \ave{\mc N(X),\mc N(X)}_{\mc N(\rho^{\otimes n})}.
\end{split}
\]
We also refer to the coarse-grained metric as the bilinear map sending $X$ and $Y$ to 
\[
\ave{\mc N(X),\mc N(Y)}_{\mc N(\rho^{\otimes n})} = \tr(\mc N(X)\Omega_{\mc N(\rho)}^{-1}\mc N(Y)).
\]

Let $A \in \mc T_\Sigma$.
Writing $\rho_\Sigma = \prod_{i \in \Sigma} \rho^{(i)}$, and $\mc N_\Sigma$ for $\mc N$ acting only on the sites $\Sigma$, we have
\begin{equation}
\mc N \Omega_{\rho^{\otimes n}}(A) = \mc N_\Sigma(\Omega_{\rho_\Sigma}(A)) \prod_{j \notin \Sigma} \mc D(\rho^{(j)}).
\end{equation}
We also observe that, in general for any states $\rho_1$, $\rho_2$, and any operator $B$,
\begin{equation}
\Omega_{\rho_1 \otimes \rho_2}(B \otimes \one) = \Omega_{\rho_1}(B) \otimes \rho_2,
\end{equation}
from which we deduce that
\(
\Omega_{\rho_1 \otimes \rho_2}^{-1}(B \otimes \rho_2) = \Omega_{\rho_1}^{-1}(B) \otimes \one.
\)
Therefore,
\begin{equation}
\label{adsislocal}
\begin{split}
\Omega^{-1}_{\mc N(\rho^{\otimes n})} \mc N \Omega_{\rho^{\otimes n}}(A) 
= \Omega^{-1}_{\mc N_\Sigma(\rho_\Sigma)} \mc N_\Sigma \Omega_{\rho_\Sigma}(A),
\end{split}
\end{equation}
which acts nontrivially solely on sites in $\Sigma$.

It follows that
\(
\|\Omega_{\rho}(A)\|_{\mc N} = \|X\|_{\mc N_{\Sigma}}
\)
where $X = \Omega_{\rho_\Sigma}(A)$ is interpreted as an operator defined solely on the Hilbert space corresponding to the sites $\Sigma$. 
Moreover, if $k := |\Sigma|$, then
\begin{equation}
\mc N_\Sigma \Omega_{\rho_\Sigma}(A) = y^{-k} \Omega_{\rho_\Sigma}(A)
\end{equation}
and hence, 
\begin{equation}
\begin{split}
\|X\|_{\mc N_{\Sigma}}^2 
&= y^{-2k} \tr(X \Omega_{\mc N_\Sigma({\rho_\Sigma})}^{-1} (X))\\
& \le y^{-2k} \tr( X \mc N_\Sigma({\rho_\Sigma})^{-\frac 1 2} X \mc N_\Sigma({\rho_\Sigma})^{-\frac 1 2} )\\
& \le y^{-2k} \tr( X^2 \mc N_\Sigma({\rho_\Sigma})^{-1}  )  \\
& \le \frac{d^k y^{-2k}  }{(1-y^{-1})^k} \tr( X^2 ).\\
\end{split}
\end{equation}
The first inequality follows from the fact that the metric we use is the smallest of the (normalised) contractive metrics, which includes the $\chi^2$ metric~\cite{petz96,petz96b}.
The second inequality is obtained using the Cauchy-Schwarz inequality on the Hilbert-Schmidt inner product. 
The last inequality follows from the fact that 
\begin{equation}
\mc N_\Sigma(\rho_{\Sigma}) \ge (1 - y^{-1})^k \tfrac{1}{d^k} \,\one.
\end{equation}
This can be seen by considering a basis diagonalizing $\rho$. 
Moreover,  
\begin{equation}
\begin{split}
\tr(X^2) &= \tfrac 1 2 \tr(\rho_\Sigma^2 A^2) + \tfrac 1 2 \tr(\rho_\Sigma A \rho_\Sigma A)  \\
&\le \tr(\rho_\Sigma^2 A^2) \le \tr(\rho_\Sigma A^2)  =\|X\|^2,
\end{split}
\end{equation} 
where we used again the Cauchy-Schwarz inequality to obtain $\tr((A \rho_{\Sigma})^\dagger \rho_{\Sigma} A) \le \tr(\rho_{\Sigma}^2 A^2)$.
Therefore, we have shown that if $X = \Omega_\rho(A)$, $A \in \mc T_{\Sigma}$, then
\begin{equation}
\label{depobound}
\|X\|_{\mc N}^2 \le \eta_k^2 \| X \|^2,
\end{equation}
where $k = |\Sigma|$ and
\begin{equation}
\eta_k^2 = \frac{d^k y^{-2k} }{(1-y^{-1})^k}.
\end{equation}

Now let us extend this to all operators $A \in \mc T_{k-1}^\perp$. We have $A = \sum_{|\Sigma| \ge k} A_\Sigma$, where $A_\Sigma \in \mc T_\Sigma$, $A_\Sigma^\dagger = A_\Sigma$.
Using Eq.~\eqref{adsislocal}, one can check that $A_\Sigma$ and $A_{\Sigma'}$ for $\Sigma \neq \Sigma'$ are not only orthogonal in the original metric, but also in the coarse-grained metric. Therefore, writing $X_\Sigma = \Omega_\rho(A_\Sigma)$, we have
\begin{equation}
\begin{split} 
\|X\|_{\mc N}^2 &= \sum_{|\Sigma|\ge k} \|X_\Sigma\|_{\mc N}^2
\le \sum_{|\Sigma|\ge k} \eta_{|\Sigma|}^2 \|X_\Sigma\|^2\\
&\le \left[{ \max_{|\Sigma| \ge k} \eta_{|\Sigma|}^2 }\right] \sum_{|\Sigma| \ge k} \|X_\Sigma\|^2 =  \max_{|\Sigma| \ge k} \eta_{|\Sigma|}^2  \|X\|^2.
\end{split}
\end{equation} 

In order to conclude, we need to determine for which minimal value of $y$ the function $\eta_k$ is decreasing as a function $k$: this is $y(y-1) > d$. Under this condition we have $\max_{|\Sigma| \ge k} \eta_{|\Sigma|}^2 = \eta_k^2$. 
\qed

\end{document}